\documentclass[twocolumn,times]{aastex62}
\usepackage{amsmath}
\usepackage{amssymb}

\newcommand{\fargo}{{\sc \tt FARGO} }
\newcommand{\bfrac}[3]{\left(\frac{#1}{#2}\right)^{#3}}

\newcommand{\dpar}[2]{\frac{\partial#1}{\partial#2}}
\newcommand{\diff}[2]{\frac{{\rm d}#1}{{\rm d} #2}}

\newcommand{\sigmagas}{\Sigma_{\rm gas}}
\newcommand{\sigmadust}{\Sigma_{\rm dust}}
\newcommand{\tstop}{t_{\rm stop}}
\newcommand{\dsize}{s_{\rm d}}

\newcommand{\dsizeo}{s_{\rm d,0}}
\newcommand{\rhop}{\rho_p}
\newcommand{\Omegak}{\Omega_{\rm K}}
\newcommand{\Omegakp}{\Omega_{\rm K,p}}

\newcommand{\st}{St}

\newcommand{\hg}{h_{\rm g}}
\newcommand{\rp}{R_{\rm p}}
\newcommand{\mpl}{M_{\rm p}}
\newcommand{\mstar}{M_{\ast}}
\newcommand{\unitr}{R_0}

\newcommand{\au}{\ \rm{AU}}

\newcommand{\hgp}{h_{\rm g,p}}

\newcommand{\sigmaunit}{\mbox{ g/cm}^2}

\newcommand{\smap}{a_{\rm p}}

\newcommand{\gammap}{\Gamma}

\newcommand{\gammapgas}{\Gamma_{\rm gas}}
\newcommand{\gammapdust}{\Gamma_{\rm dust}}
\newcommand{\rdout}{R_{\rm out}}
\newcommand{\rdin}{R_{\rm in}}
\newcommand{\gammafb}{\Gamma_{\rm fb}}
\newcommand{\gammanofb}{\Gamma_{\rm nofb}}

\received{\today}
\revised{\today}
\accepted{\today}
\submitjournal{ApJL}

\shorttitle{Termination of an inward migration of a gap-opening planet triggered by dust feedback}
\shortauthors{K.D. Kanagawa}

\begin{document}

\title{Termination of an inward migration of a gap-opening planet triggered by dust feedback}

\correspondingauthor{Kazuhiro D. Kanagawa}
\email{kazuhiro.kanagawa@utap.phys.s.u-tokyo.ac.jp}

\author[0000-0001-7235-2417]{Kazuhiro D. Kanagawa}
\affiliation{Research Center for the Early Universe, Graduate School of Science, University of Tokyo, Hongo, Bunkyo-ku, Tokyo 113-0033, Japan}



\begin{abstract}
The planet migration due to the disk--planet interaction is one of the most important processes to determine the architecture of planetary systems.
A sufficiently massive planet forms a density gap and migrates together with the gap.
By carrying out two-dimensional and two-fluid (gas and dust grains) hydrodynamic simulations, we investigated the effects of the dust feedback on the migration of the gap-opening planet, which was not considered in previous studies.
We found that the gas surface density at the outer edge of the gap becomes smaller due to the dust feedback, and thus the torque exerted from the outer disk decreases.
This mechanism becomes effective as the gap becomes wider and deeper.
In particular, when the mass of the planet is Jupiter-size and turbulent viscosity is $\alpha = 3\times 10^{-4}$, the planet can migrate outward due to the reduction of the torque exerted from the outer disk.
Even for a smaller planet, the migration becomes significantly slow down.
This termination of the inward migration triggered by the dust feedback may explain why ring and gap structures can be frequently observed within the protoplanetary disks.
\end{abstract}



\section{Introduction} \label{sec:intro}
A planet within a protoplanetary disk interacts with a surrounding disk gas and migrates in radial direction \citep[e.g.,][]{Goldreich_Tremaine1980,Lin_Papaloizou1979}.
A sufficiently massive planet forms a density gap as a consequence of disk--planet interaction, and it migrates together with the gap \citep[e.g.,][]{Lin_Papaloizou1986b,Ward1997}.
This planet migration continues until the gaseous disk is depleted, which determine the final architecture of planetary systems.
Therefore, the planet migration is one of the most important processes in the formation of the planetary system.

The radial migration of the gap-opening planet is called type~II migration.
In the ideal case, the gap induced by the massive planet halts the gas flow across it and hence the planet migrates in the same velocity of the viscous drift of the gas \citep[e.g.,][]{Lin_Papaloizou1986b,Armitage2007}.
However, recent numerical hydrodynamic simulations have shown that the planetary gap hardly halts the gas flow across it, even if the planet is as massive as Jupiter \citep[e.g.,][]{Duffell_Haiman_MacFadyen_DOrazio_Farris2014,Durmann_Kley2015,Durmann_Kley2017,Kanagawa_Tanaka_Szuszkiewicz2018,Robert_Crida_Lega_Meheut_Morbidelli2018}.
By carrying out two-dimensional hydrodynamic simulations in a broad parameter range, \cite{Kanagawa_Tanaka_Szuszkiewicz2018} have found that the torque exerted on the planet is roughly proportional to the surface density at the bottom of the gap, though strictly speaking, the torque is also exerted from the bottom and the edge of the gap.
This result indicates that the migration speed of the gap-opening planet is sensitive to the gas structure of the gap.

At the outer edge of the gap, relatively large dust grains, so-called ``pebbles'', are highly pilled-up, because these dust grains cannot cross the gap region \citep[e.g.,][]{paardekooper2004,Muto_Inutsuka2009b,Zhu2012,Dong_Zhu_Whitney2015,Pinilla_Ovelar_Ataiee_Benisty_Birnstiel_Dishoeck_Min2015,Weber2018,Meru_Rosotti_Booth_Nazari_Clarke2018,Kanagawa_Muto_Okuzumi_Taki_Shibaike2018}.
The dust ring formed at the outer edge of the gap might be associated with bright rings in the protoplanetary disks observed by e.g., ALMA \citep[e.g.,][]{ALMA_HLTau2015,Nomura_etal2016,Tsukagoshi2016,Akiyama2016,van_der_Plas_etal2017,Fedele_etal2017,Dong2018,PDS70_Long2018,Taurus_Long2018,DSHARP1}.
The dust grains lose angular momentum due to friction with the surronding disk gas and fall down to the central star.
Simultaneously the disk gas receives the same amount of angular momentum that the dust grains lose, which is referred to as dust feedback \citep{Fu2014,Gonzalez2015,Gonzalez2017,Taki_Fujimoto_Ida2016,Dipierro_Laibe2017,Kanagawa_Ueda_Muto_Okuzumi2017,Weber2018,Dipierro_Laibe_Alexander_Hutchison2018,Kanagawa_Muto_Okuzumi_Taki_Shibaike2018}.
Recent gas--dust two fluid hydrodynamic simulations \citep{Weber2018,Kanagawa_Muto_Okuzumi_Taki_Shibaike2018} have found that though it is not significant, the dust feedback can modify the outer edge of the gap.
This slight modification induced by the dust feedback may significantly affect the migration speed, because it could break the balance between torques exerted from the inner and outer disks.

We investigate the effects of the dust feedback on the migration speed of the gap-opening planet, by carrying out the gas--dust two fluid and two-dimensional hydrodynamic simulations.
In Section~\ref{sec:basic_eq}, our numerical method and setup are described.
The results of our hydrodynamic simulations are shown in Section~\ref{sec:results}.
In Section~\ref{sec:discussion}, we summarize and discuss our results.

\section{Numerical method} \label{sec:basic_eq}
We implemented the dust component into the hydrodynamic code \fargo \citep{Masset2000}, and numerically solved the equations of motion and the continuity equations for the gas and the dust grains \citep{Kanagawa_Ueda_Muto_Okuzumi2017,Kanagawa_Muto_Okuzumi_Taki_Shibaike2018}.
We include the gas-dust drag force in the equation of motion of the gas, as well as that of the dust grains.
Due to this drag force in the equation of the motion of the gas (dust feedback), the gas is affected by the dust grains, similar to that the dust grains drift because of the drag force with the gas.
In the present study, we take into account the planet migration, which was not considered by the previous studies.
We choose a two-dimensional cylindrical coordinate system ($R,\phi$), and its origin is located at the position of the central star.

\subsection{Planetary migration} \label{subsec:migration}
Here we briefly describe the method to implement the migration of the planet.
We assume that the eccentricity of the planet is much smaller than unity (this assumption is valid in our simulations shown in this letter).
In this case, the equation of motion of the planet is written by
\begin{align}
\frac{1}{\smap} \diff{\smap}{t} &= \frac{2\gammap}{\mpl \smap^2 \Omegakp},
\label{eq:timevar_semajoraxis2}
\end{align}
where $\mpl$, $\smap$, and $\Omegakp$ denote the mass, semi-major axis, and the Keplerian angular velocity of the planet, respectively.
In the following, the subscript ``${\rm p}$'' indicates a physical quantity at the orbital radius of the planet $R=\rp$.
Since we consider a small eccentricity, $\rp \simeq \smap$.
The torque exerted on the planet $\gammap$ is given by the sum of these exerted from the gas disk and the dust disk, as
\begin{align}
\gammap &= \gammapgas + \gammapdust,
\label{eq:gammap}
\end{align}
where $\gammapgas$ and $\gammapdust$ are given by \citep{Kley_Nelson2012}
\begin{align}
\gammapgas &= \int^{\rdout}_{\rdin}RdR \int^{2\pi}_{0} d\phi \sigmagas \dpar{\psi}{\phi}, \label{eq:gammagas}\\
\gammapdust &= \int^{\rdout}_{\rdin}RdR \int^{2\pi}_{0} d\phi \sigmadust \dpar{\psi}{\phi}, \label{eq:gammadust}
\end{align}
where $\psi$, $\rdin$ and $\rdout$ denote the gravitational potential, and the radii of the inner and outer edge of the disk, respectively.
The surface densities of the gas and the dust grains are represented by $\sigmagas$ and $\sigmadust$, respectively.
We introduce a characteristic torque $\Gamma_0 $ as
\begin{align}
\Gamma_0 (R) &= \bfrac{\mpl}{\mstar}{2} \bfrac{\hg}{R}{-2} \sigmagas{}_{\rm , un} R^4 \Omegak^2,
\label{eq:gamma0}
\end{align}
where $\Omegak$ and $\sigmagas{}_{\rm , un}$ denote the Keplerian angular velocity and the gas surface density of the unperturbed disk, respectively, which are the functions of $R$.
Note that $\Gamma_0$ roughly corresponds to the absolute value of the torque expected by the Type~I regime.

For convenience, we define a cumulative torque from the gas and the dust grains as
\begin{align}
T_{\rm gas}(R) &= \int^{\rdout}_{R} \frac{d\Gamma_{\rm gas}}{dR'}dR'\\
T_{\rm dust}(R) &= \int^{\rdout}_{R} \frac{d\Gamma_{\rm dust}}{dR'}dR'.
\end{align}
Thus, the cumulative torque exerted from the gas and the dust grains are given by $T(R)=T_{\rm gas}(R)+T_{\rm dust}(R)$.

\subsection{Setup of numerical simulations} \label{subsec:method}
We use the initial orbital radius of the planet $\unitr$ and the mass of the central star $\mstar$ as the unit of length and mass, respectively.
Hence, the surface density is normalized by $\mstar/\unitr^2$.
We also use the unit of time $t_0$ which is defined by $2\pi/\Omegak (\unitr)$.
The computational domain runs from $R/R_0=0.3$ to $R/R_0=3.2$ with 512 radial zones (equally spaced in logarithmic space) and 1024 azimuthal zones (equal spaced).
The initial surface density of the gas is set as $\sigmagas{}_{\rm , un}=\Sigma_0 \left(R/\unitr \right)^{-1}$ with $\Sigma_0=10^{-4}$.
For the dust grains, the initial surface density is set for the dust-to-gas mass ratio to be 0.01.
The disk aspect ratio are assumed to be $\hg/R = H_0 (R/\unitr)^{1/4}$ which is assumed to be independent of time, where $H_0$ is set to be $0.05$ in this letter and $\hg$ is the gas scale height.
The initial angular velocity of the gas is given by $\Omegak \sqrt{1-\eta}$ and $\eta=-1/2(\hg/R)^2 \partial (\ln P)/\partial (\ln R)$.
For the dust grains, the initial angular velocity is given by $\Omegak$.
The initial radial velocities of the gas and the dust grains are set to be $-3\nu/R$ and zero, respectively, where $\nu$ is the kinetic viscosity.
We use the $\alpha$-prescription of \cite{Shakura_Sunyaev1973}, and hence $\nu = \alpha \hg^2 \Omegak$.
We also assume that $\alpha$ is constant throughout the computational domain.
The smoothing parameter of the gravitational potential is set to be $0.6\hgp$.
We set wave killing zones with the width of $0.1R_0$ for the gas at the inner and outer boundaries, as described in \cite{Kanagawa_Tanaka_Szuszkiewicz2018}.

We assume constant-size dust grains.
The size of the dust grains $\dsize$ is set to be $0.1\dsizeo$, where $\dsizeo$ is defined by
\begin{align}
\dsizeo &=\frac{2\Sigma_0}{\pi \rhop} \nonumber \\
        &= 57.1 \bfrac{\Sigma_0}{8.9 \sigmaunit}{} \bfrac{\rhop}{1\mbox{ g/cm}^3}{-1} \mbox{ cm},
\end{align}
where $\rhop$ denotes the internal density of the dust grains.
Note that when $\mstar=1M_{\odot}$ and $\unitr=10 \au$, the dimensionless surface density $\Sigma = 10^{-4}$ corresponds to 8.9 $\mbox{g/cm}^2$.
Adopting the Epstein regime, one can write the stopping time of dust grains for the two-dimensional disk $\tstop$ as
\begin{align}
\tstop &= \frac{\pi \dsize \rhop}{2\sigmagas \Omegak}.
\label{eq:tstop}
\end{align}
The Stokes number of dust grains, $\st$, is defined by
\begin{align}
\st &=\tstop \Omegak .
\label{eq:st}
\end{align}

\section{Results} \label{sec:results}
\subsection{Halting inward planetary migration triggered by the dust feedback} \label{subsec:example_cases}
First we present the result of our simulations in the case of a Jupiter-mass planet ($\mpl/\mstar = 1\times 10^{-3}$) with $\alpha=3\times 10^{-4}$ and $H_0=0.05$.
\begin{figure}
  \begin{center}
  \resizebox{0.49\textwidth}{!}{\includegraphics{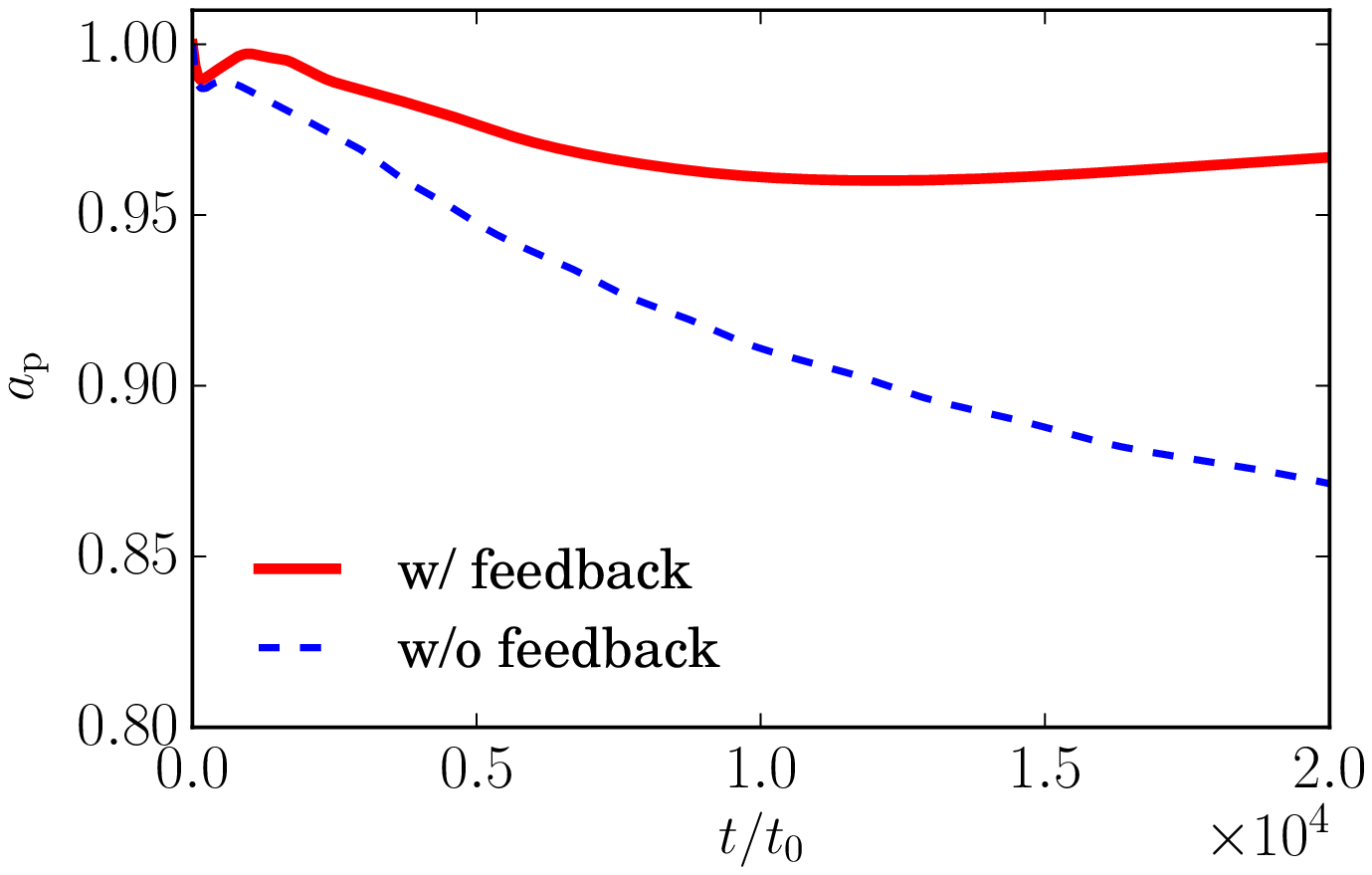}}\\
  \resizebox{0.49\textwidth}{!}{\includegraphics{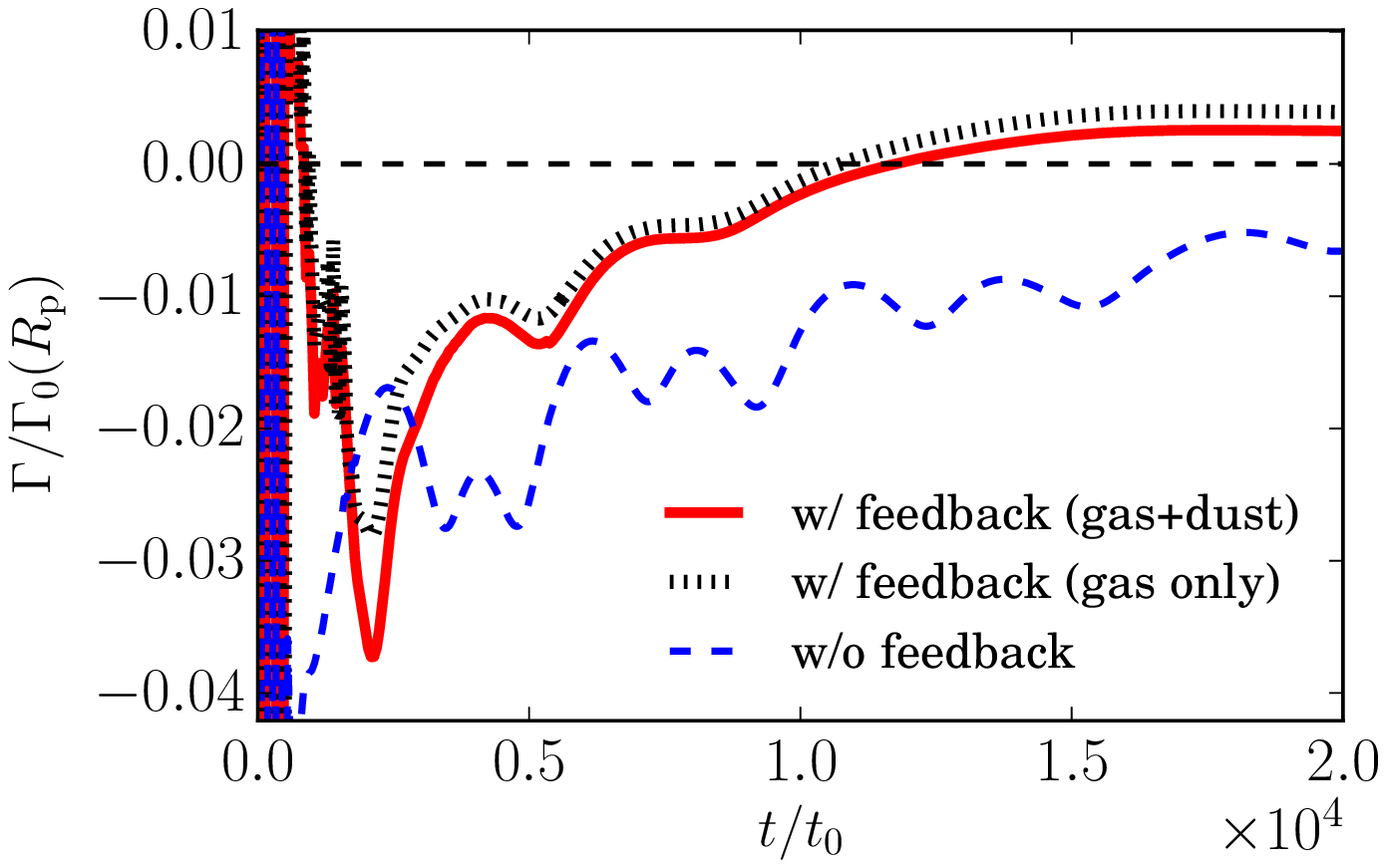}}\\
  \resizebox{0.49\textwidth}{!}{\includegraphics{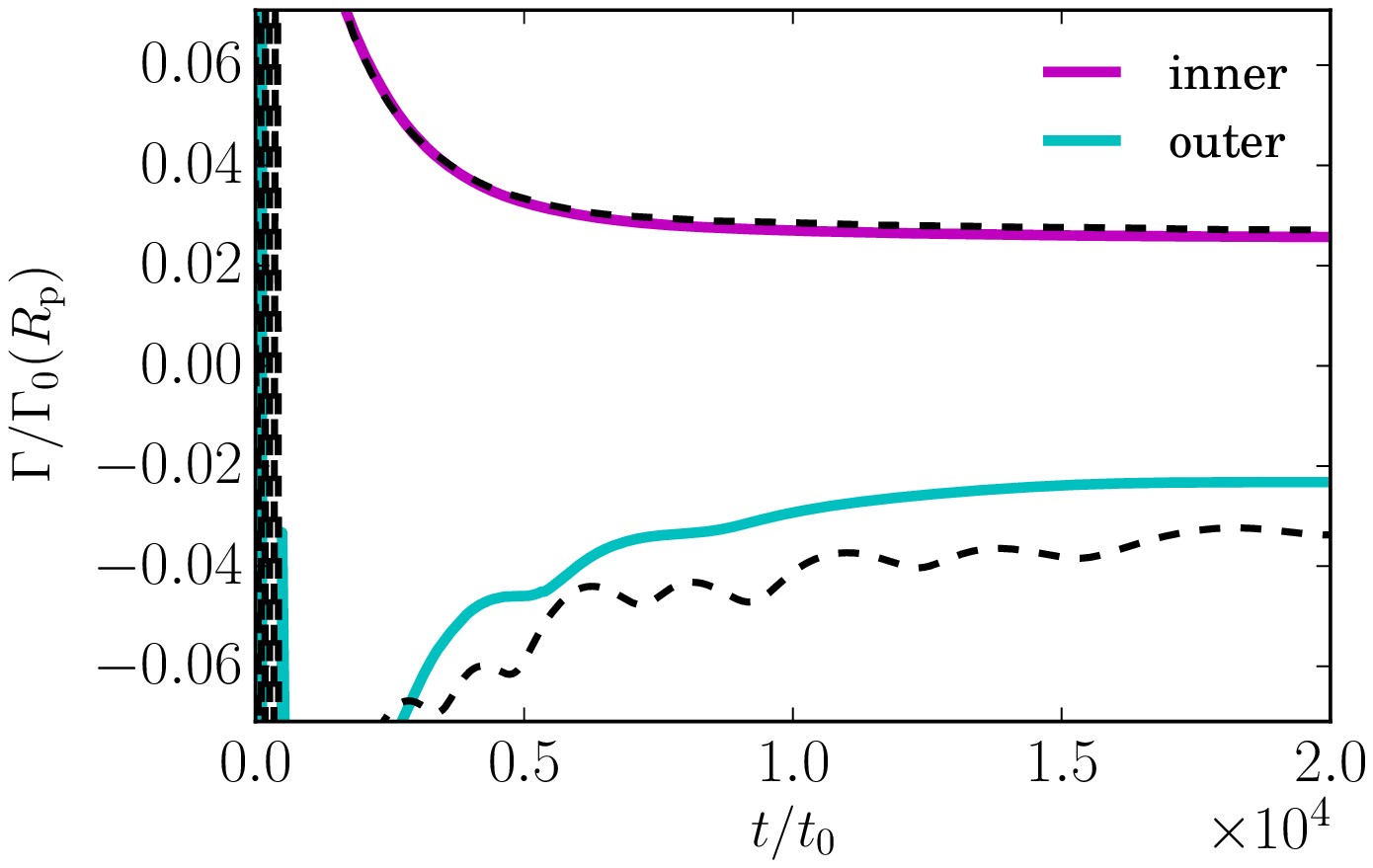}}
  \caption{
  Evolutions of the semimajor axis of the planet (top), total torque eserted on the planet (middle), and the torques exerted from the inner and outer disks (bottom), when $\mpl/\mstar=1\times 10^{-3}$, $H_0=0.05$ and $\alpha=3\times 10^{-4}$.
  \label{fig:evo_a3e-4_h5e-2_q1e-3}
  }
  \end{center}
\end{figure}
The top panel of Figure~\ref{fig:evo_a3e-4_h5e-2_q1e-3} compares the evolution of the semi-major axis of the planet in the case with and without the dust feedback.
It is obvious that the migration rate is significantly lower when the dust feedback is included.
In particular, after $t \simeq 12000 \ t_0$, the inward migration is terminated and the planet migrates outward slowly.

The middle and bottom panels of Figure~\ref{fig:evo_a3e-4_h5e-2_q1e-3} show a total torque exerted on the planet and the torques exerted from the inner and outer disk, respectively.
The absolute value of torque exerted on the planet in the case with the dust feedback is significantly smaller than that in the case without the dust feedback.
After $t\simeq 12000 t_0$, the total torque becomes positive and hence the planet moves outward in the case with the dust feedback.
As can be seen from the bottom panel of Figure~\ref{fig:evo_a3e-4_h5e-2_q1e-3}, only the torque exerted from the outer disk in the case with the dust feedback decreases as compared to that in the case without the dust feedback.
Since the negative torque exerted from the outer disk descreases, the inward migration of the planet is halted and finally the planet migrates outward, in the case with the dust feedback.

\begin{figure}
  \begin{center}
  \resizebox{0.49\textwidth}{!}{\includegraphics{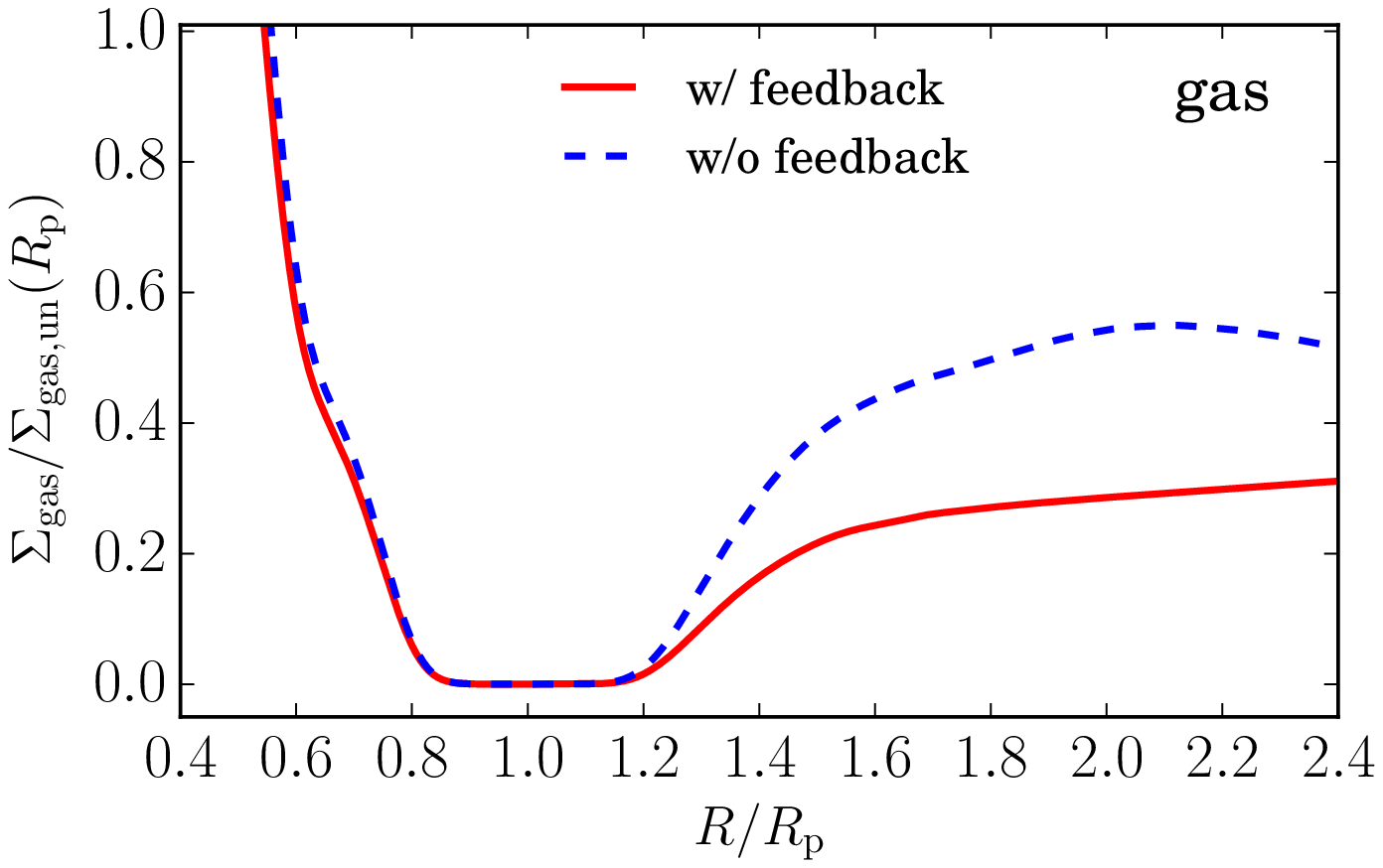}} \\
  \resizebox{0.49\textwidth}{!}{\includegraphics{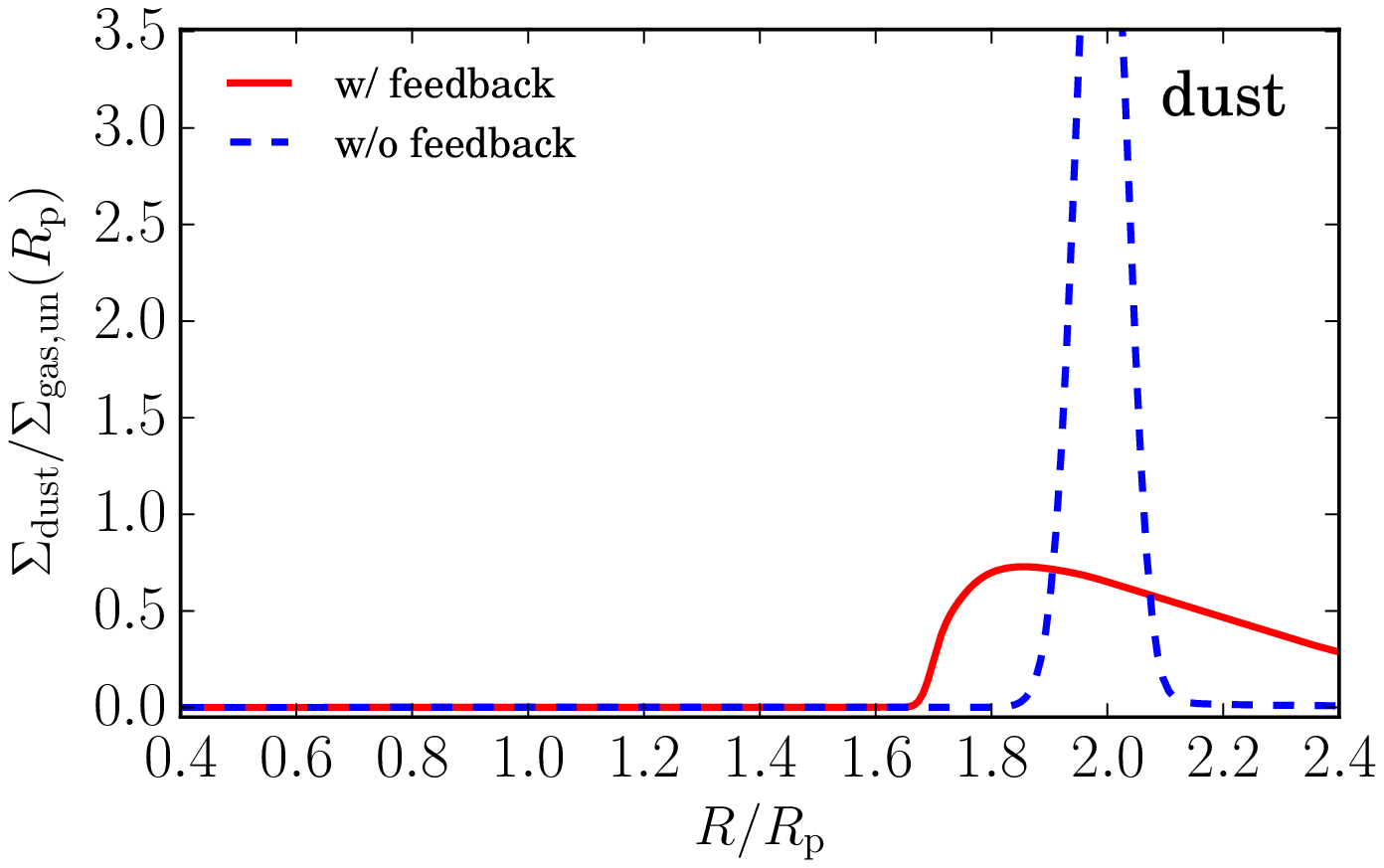}}\\
  \resizebox{0.49\textwidth}{!}{\includegraphics{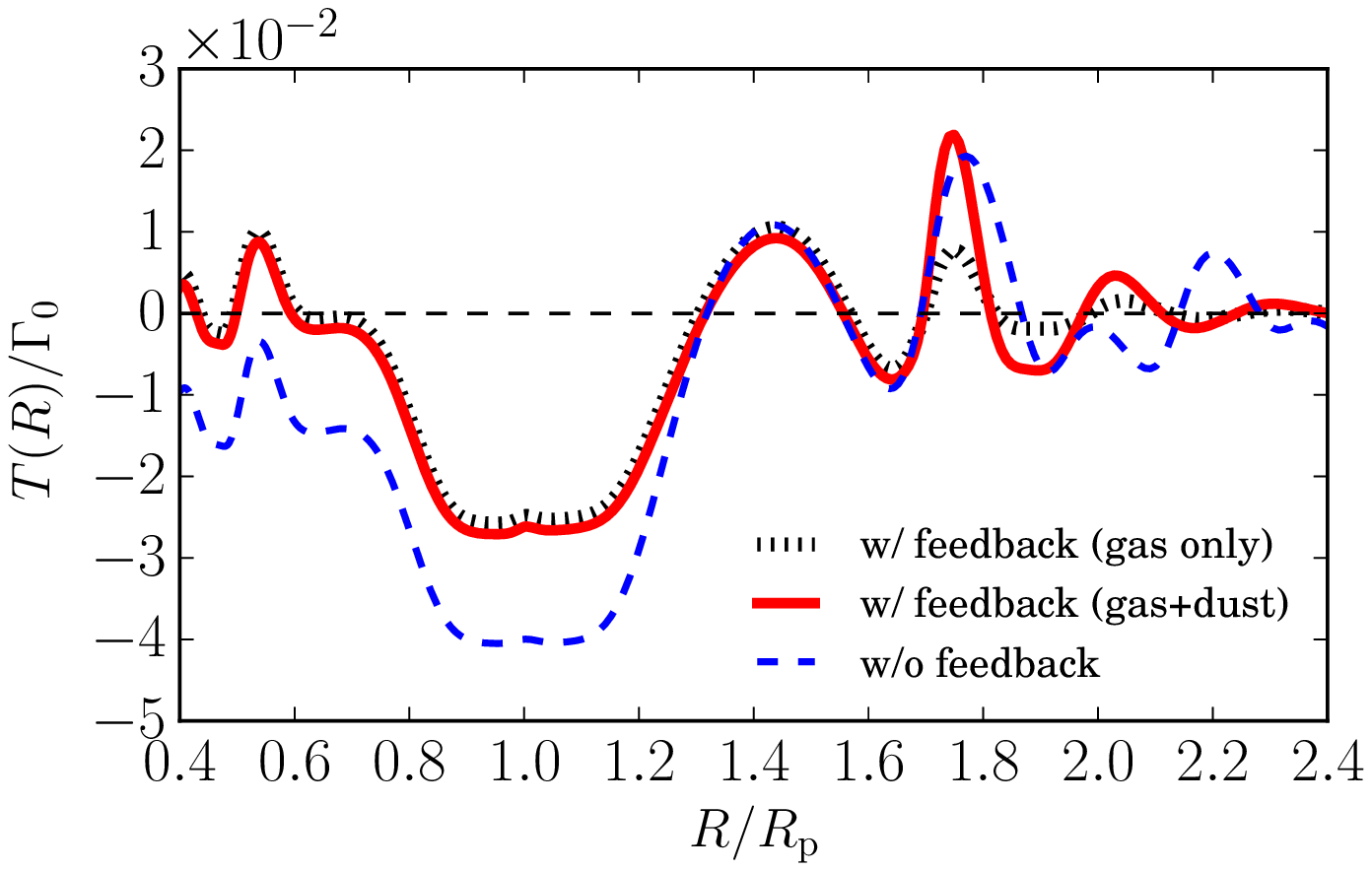}}
  \caption{
  Azimuthally averaged surface densities of gas (top), dust grains (middle), and cumulative torque (bottom) at $t=12000\ t_0$ in the case in which the dust feedback is turned on and that in the case in which the dust feedback is turned off, when $\mpl/\mstar=10^{-3}$, $H_0=0.05$ and $\alpha=3\times 10^{-4}$.
  \label{fig:avgdens_a3e-4_h5e-2_q1e-3}
  }
  \end{center}
\end{figure}
The top and middle panels of Figure~\ref{fig:avgdens_a3e-4_h5e-2_q1e-3} show the azimuthal averaged surface densities of the gas and the dust grains at $t=12000\ t_0$, in the cases with and without the dust feedback.
The dust feedback makes the gradient of the gas surface density small (in other words, the value of $\eta$ becomes small) in the outer edge of the gap, as shown by \cite{Kanagawa_Muto_Okuzumi_Taki_Shibaike2018}; (see also \cite{Taki_Fujimoto_Ida2016}).
As a result, the surface density of the gas in the outer disk of $R>1.3 \rp$ becomes smaller than  that without the dust feedback.
Since the dust drift velocity is much faster than that of the planet, the dust grains drifting from the outer disk are trapped into the gas and they form the dust ring at the outer edge.
In the inner disk, most of the dust grains drift to the central star.
The distribution of the dust grains against the planet's orbit is a consequence of the dust trap by the gap and the fast dust drift.

In the bottom panel of Figure~\ref{fig:avgdens_a3e-4_h5e-2_q1e-3}, we show the cumulative torques.
The torque is exerted from the bottom to the edge of the gap, as \cite{Kanagawa_Tanaka_Szuszkiewicz2018} stated, and the torque in the case without the dust feedback is consistent to that given by the formula of \cite{Kanagawa_Tanaka_Szuszkiewicz2018}.
In the case with the dust feedback, the slope of the cumulative torque is significantly less steeper than that in the case with the dust feedback because of the depletion of the outer disk due to the dust feedback, though they are similar to each other in other regions.
Therefore, the torque exerted from the outer disk is smaller and finally the outward migration is caused in the case with the dust feedback, as seen in Figure~\ref{fig:evo_a3e-4_h5e-2_q1e-3}.

\subsection{Parameter study} \label{subsec:high_vis_case}
In order to see how sensitive the effects of the dust feedback to planet masses ($\mpl$) and viscosities ($\alpha$), we performed a series of simulations.
\begin{figure}
  \begin{center}
  \resizebox{0.49\textwidth}{!}{\includegraphics{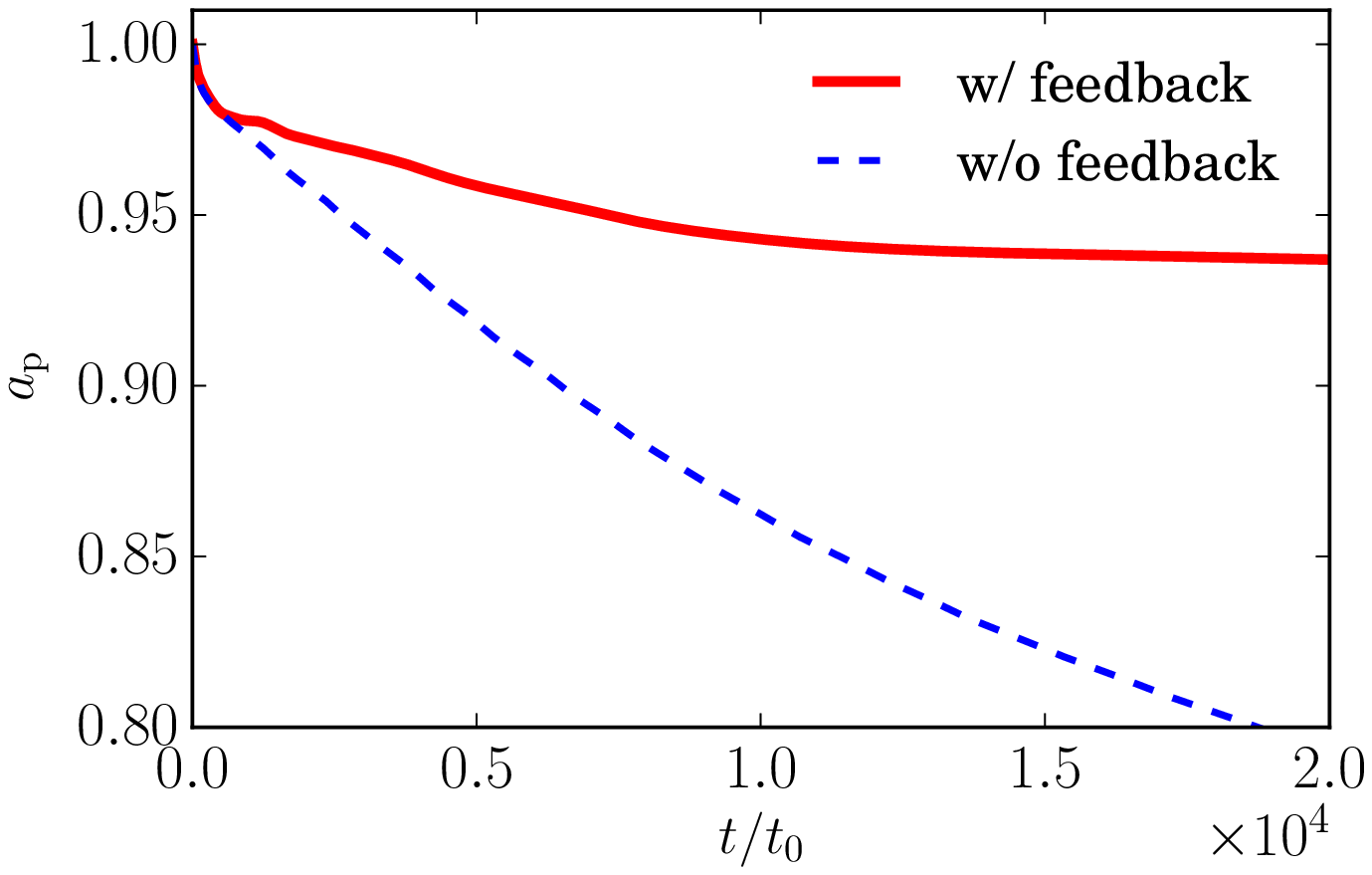}}\\
  \resizebox{0.49\textwidth}{!}{\includegraphics{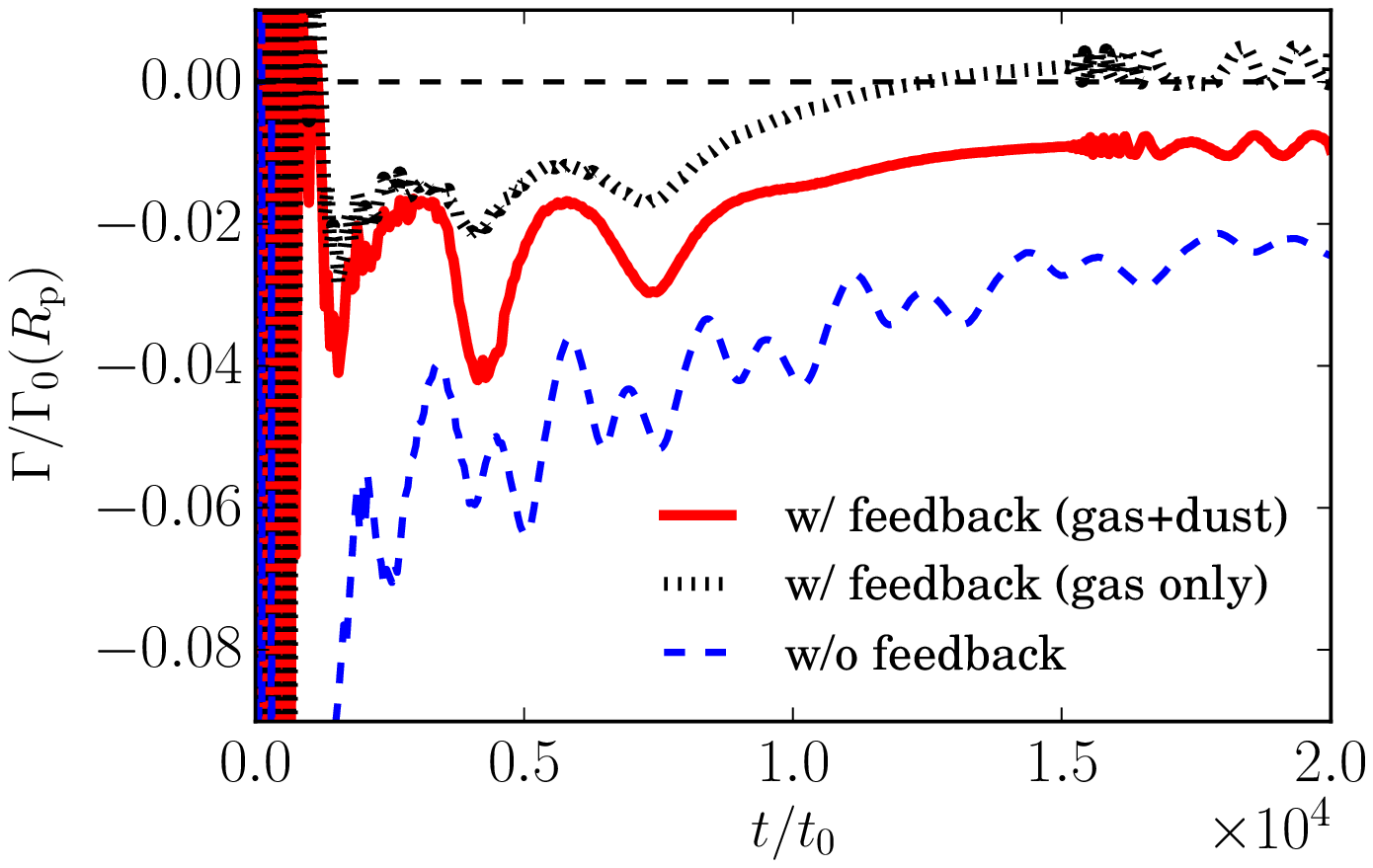}}\\
  \resizebox{0.49\textwidth}{!}{\includegraphics{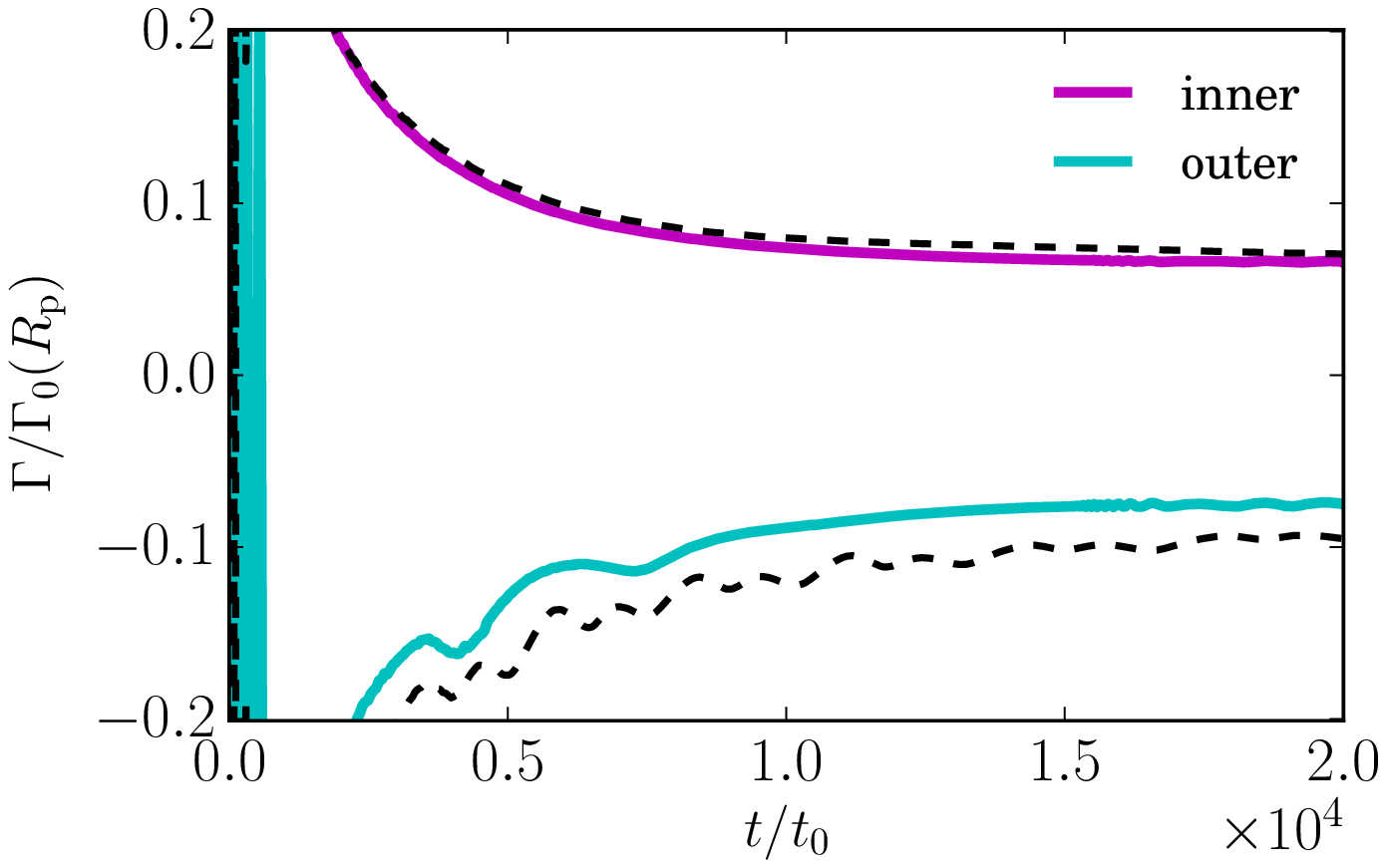}}
  \caption{
  The same as Figure~\ref{fig:evo_a3e-4_h5e-2_q1e-3}, but $\mpl/\mstar=5\times 10^{-4}$.
  \label{fig:evo_a3e-4_h5e-2_q5e-4}
  }
  \end{center}
\end{figure}
First, in Figure~\ref{fig:evo_a3e-4_h5e-2_q5e-4}, we show the case of $\mpl/\mstar=5\times 10^{-4}$, but $H_0$ and $\alpha$ are the same as Figure~\ref{fig:evo_a3e-4_h5e-2_q1e-3}.
Similar to the case of Figure~\ref{fig:evo_a3e-4_h5e-2_q1e-3}, the value of $|\Gamma/\Gamma_0|$ decreases because the torque exerted from the outer disk decreases, in the case with the dust feedback.
As a result, the inward migration in the case with the dust feedback is much slower than that in the case without the dust feedback.
However, in this case, since the gap is narrow and the dust ring is formed at closer from the planet, as compared to the case shown in Figure~\ref{fig:evo_a3e-4_h5e-2_q1e-3}, the planet feels the negative torque exerted from the dust ring significantly.
Becuase of it, the inward migration of the planet continues, though the torque exerted from the gas is positive after $t=12000 \ t_0$.

\begin{figure}
  \begin{center}
  \resizebox{0.49\textwidth}{!}{\includegraphics{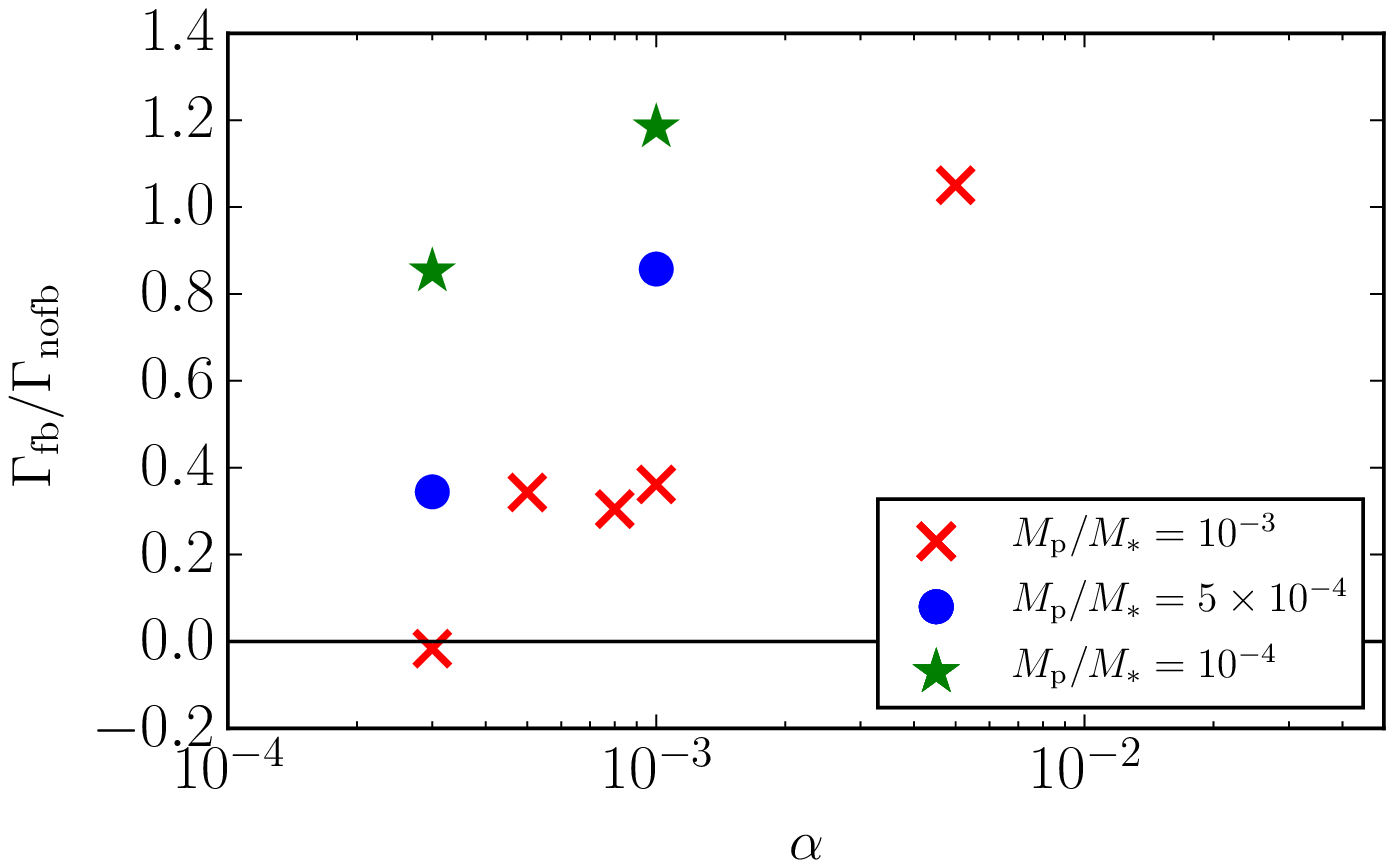}} \\
  \resizebox{0.49\textwidth}{!}{\includegraphics{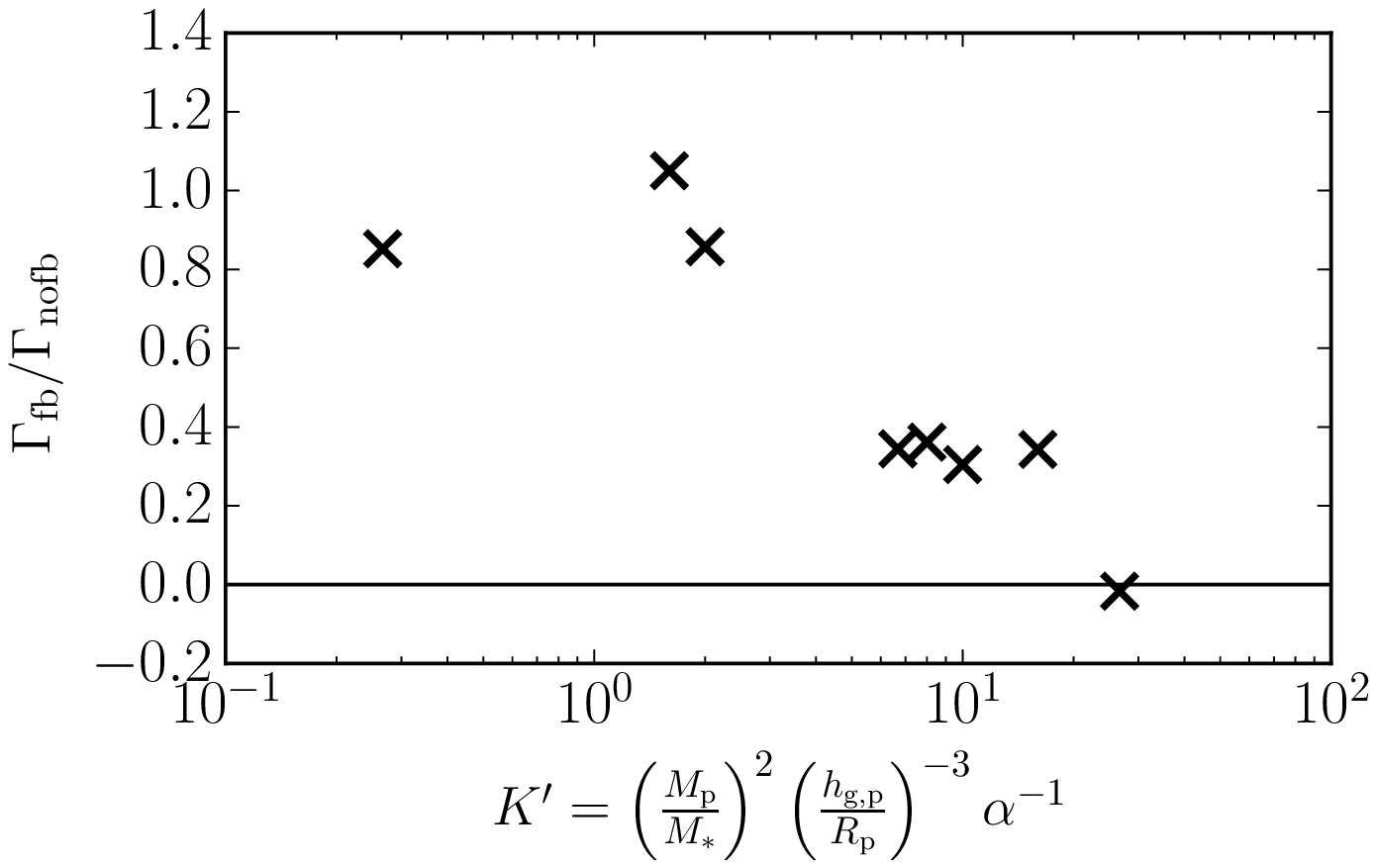}}
  \caption{Ratios between the torques in the case in which the dust feedback is turned on and in the case in which the feedback is turned off (reduction rate), against the values of viscous $\alpha$ (top) and against $K'$ (bottom).
  \label{fig:param_study}
  }
  \end{center}
\end{figure}
The top panel of Figure~\ref{fig:param_study} shows the dependence of the viscosity on the ratio of the torques with and without the feedback the dust feedback, $\gammafb$ and $\gammanofb$, for various $\mpl/\mstar$ with $H_0=0.05$.
Specifically, the value of torques in Figure~\ref{fig:param_study} corresponds to the average over $11500\ t_0 < t < 12000\ t_0$ \footnote{Only in the case of $\mpl/\mstar=10^{-4}$ and $\alpha=10^{-3}$, we show the torque at $\sim 8000\ t_0$, because the planet reaches too close to the inner boundary at $\sim 12000\ t_0$.}.
Note that though we took the torque at $\sim 12000\ t_0$, the torque at that time may not exactly correspond to the stationary value, as in the case shown in Figure~\ref{fig:evo_a3e-4_h5e-2_q1e-3}.
Hence, the figure shows just a rough trend.

The top panel of Figure~\ref{fig:param_study} indicates that $\gammafb/\gammanofb$ decreases as the decrease of $\alpha$, because the gap becomes wider with an decrease of $\alpha$.
The bottom panel of Figure~\ref{fig:param_study} shows this tendency more clear.
The bottom panel is the same as the top panel, but the horizontal axis of the bottom panel is the dimensionless parameter $K'=(\mpl/\mstar)^2(\hgp/\rp)^{-3} \alpha^{-1}$, which is related to the gap width of the gas in stationary state as $\Delta_{\rm gap} = 0.41 K'^{1/4}$ \citep{Kanagawa2016a}.
Larger value of $K'$ indicates wider gap.
As can be seen in the bottom panel of Figure~\ref{fig:param_study}, the ratio of $\gammafb$ to $\gammanofb$ decreases as $K'$ increases (the gap is wider), when $K'\gtrsim 1$.

\section{Summary and Discussion} \label{sec:discussion}
In this letter, we have shown that the dust feedback reduces the surface density at the outer edge of the gap and the torque exerted on the planet from the outer disk.
As a result, the radial migration of the planet can be significantly slow .
In particular, when the viscosity is low, namely $\alpha=3\times 10^{-4}$, the inward migration is halted and the Jupiter-mass planet moves outward slowly.

\begin{figure}
  \begin{center}
  \resizebox{0.49\textwidth}{!}{\includegraphics{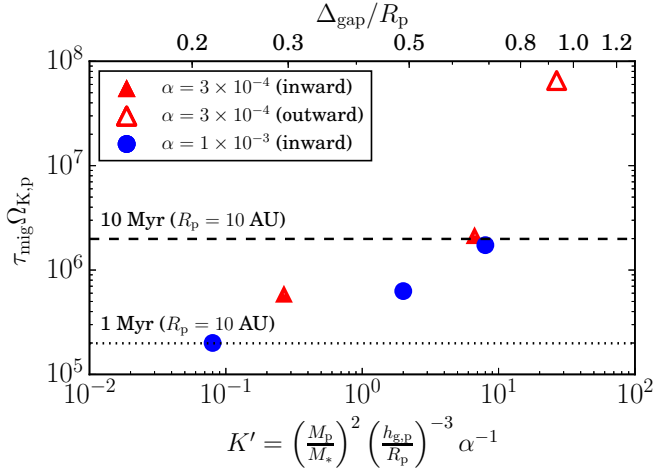}}
  \caption{
  Migration timescales obtained by our simulations against $K'$ that is related to the width of the gap in the case of the relatively light disk ($M_{\rm disk} \simeq 0.006M_{\odot}$).
  Filled symbols indicate the timescale of the inward migration and a open symbol indicates the timescale of the outward migration.
  The width of the gap estimated by $0.41K'^{1/4}$ is indicated in the upper horizontal axis.
  The dotted and dashed horizontal lines correspond to $\tau_{\rm mig}=1 \mbox{ Myr}$ and $\tau_{\rm mig} = 10 \mbox{ Myr}$ when $\rp=10\mbox{ AU}$, respectively.
  \label{fig:tmig_vs_width}
  }
  \end{center}
\end{figure}
In the rest of this letter, we discuss implications of our results for the planet formation and the observations.
The planet migration within the disk continues until the disk gas removal, typically at a few million years, which determine the final position of the planet.
Figure~\ref{fig:tmig_vs_width} shows the migration timescales, $\tau_{\rm mig} = - ({\rm d}\ln \smap/{\rm d}t)^{-1}$, calculated by our simulations against the dimensionless parameter $K'$ (which is mentioned in the previous section).
The estimated width of the gap from $K'$ is indicated in the upper horizontal axis of the figure.
This estimated width is just a reference because the simulations of \cite{Kanagawa2016a} does not consider the dust feedback, and also the dust gap can be wider than the gap of the gas, depending on the size of the dust grains \citep[e.g.,][]{Zhu2012,deJuanOverlar2013}.
We can consider the estimated width as a rough lower limit of the gap width.

As can be seen in Figure~\ref{fig:tmig_vs_width}, the migration timescale is longer as the increase of $K'$.
In particular, when $K'\gtrsim 1$ ($\Delta_{\rm gap}/\rp \gtrsim 0.5$), the timescale of the inward migration is about $10\mbox{ Myr}$, which is a order-of-manugnitude larger than the typical disk lifetime.
Moreover, when the gap is sufficiently wide, namely, $K'=20$, the planet migrates outward as shown in Section~\ref{subsec:example_cases}.
Note that in the figure, the relative light disk is assumed as $M_{\rm disk} = 0.006M_{\odot}$.
For massive disks such as the disk of HL~Tau ($M_{\rm disk} \simeq 0.1M_{\odot}$), the migration timescale is shorter ($\times \sim 0.1 $) than that shown in Figure~\ref{fig:tmig_vs_width}.
Even for the massive disk, in the cases with the timescale of $\sim 10 \mbox{ Myr}$ shown in Figure~\ref{fig:tmig_vs_width}, the planet can survive until the removal of the disk gas.

Recent observations have revealed that a number of the protoplanetary disks have ring and gap structures.
Here we discuss the observability of the gap formed by the planet.
The timescales of the gap formation and planet migration are proportional to the Keplerian rotation period.
That is, these timescales are longer in the region with larger radii.
Hence, in the outer region, the gap shape may not reach stationary state yet and the width of the gap is narrower than the stationary width.
By considering the long timescale of the gap formation and migration, it is understandable that we can observe the narrow gap in the outer region.
However, in the inner region, the migration timescale becomes shorter as the Keplerian rotation is faster.
Hence, the gap is difficult to be observed in the inner region due to the fast inward migration, without the mechanism which slows the migration down.
Only when the planet forms a relatively wide gap, namely $\Delta_{\rm gap}/\rp \gtrsim 0.5$, the planet could survive during the disk lifetime, due to the effect of the dust feedback.
Hence, the width of the gap would be wide in the inner region, if the observed gap is formed by the planet.
Actually, ALMA observations have found such a wide gaps in the relatively inner region, for instance, in the disks of HD143006 \citep{Huang_DSHARP}, HL~Tau \citep{ALMA_HLTau2015,Kanagawa2016a}, and PSD~70 \citep{PDS70_Long2018}.
These gaps could be formed by the planet.
In the near future, ALMA will reveal the gaps within the region of $\lesssim 10 \mbox{ AU}$.
Our results imply that these gaps are as wide as $\Delta_{\rm gap}/\rp \gtrsim 0.5$.

Within the dust ring in which the dust grains are highly accumulated, the planetesimals and protoplanets can be formed via e.g., the streaming instability \citep[e.g.,][]{Youdin_Goodman2005}.
The planet migration is very slow until the dust grains are removed by the planetesimal formation.
Once the planet formation starts within the dust ring, a number of planets can be formed within the dust ring as discussed in \cite{Kanagawa_Muto_Okuzumi_Taki_Shibaike2018}.
Otherwise, a remnant of the dust ring may survive.
Such a path of the planet formation may explain the large diversity of the exoplanets revealed by observations.

The suppression and turnover of the inward migration of the gap-opening planet shown in this letter are critical for the planetary system formation and the origin of the observed gaps.
Although we investigated only the cases with $H_0=0.05$, the further survey covering a wider parameter range should be done to determine the condition of the onset of the outward migration and quantitative relation of reduced migration time and planet mass.

\acknowledgements
KDK would like to thank Takayuki Muto for an insightful discussion and constructive comments, and Professor Yasushi Suto for helpful comments.
KDK also thank the anonymous referee for thoughtful comments.
KDK was supported by JSPS Core-to-Core Program ``International Network of Planetary Sciences'' and JSPS KAKENHI grant No. 19K14779.
Numerical computations were carried out on the Cray XC50 at the Center for Computational Astrophysics, National Astronomical Observatory of Japan.


\end{document}